\RequirePackage{ifpdf}
\ifpdf 
\documentclass[pdftex]{sigma}
\else
\documentclass{sigma}
\fi

\newcommand{\gam}{{}^{\text{\tiny{($\gamma$)}}}\!}
\newcommand{\bst}{{}^{\text{\tiny{$\mathcal{B}$}}}\!\!}
\newcommand{\rot}{{}^{\text{\tiny{$\mathcal{R}$}}}\!\!}
\newcommand{\brA}{{}^{\text{\tiny{$\mathcal{B},\mathcal{R}$}}}\!\!\mathcal{A}}
\newcommand{\ub}{\underline{u}\,}
\newcommand{\vb}{\underline{v}\,}
\newcommand{\Yb}{\underline{Y}\,}
\newcommand{\SU}{\text{SU}}

\newcommand{\SL}{\text{SL}}
\newcommand{\su}{\mathfrak{su}}
\newcommand{\so}{\mathfrak{so}}

\def\f{\frac}

\begin{document}

\allowdisplaybreaks

\renewcommand{\thefootnote}{$\star$}

\renewcommand{\PaperNumber}{083}

\FirstPageHeading

\ShortArticleName{A Lorentz-Covariant Connection for Canonical Gravity}

\ArticleName{A Lorentz-Covariant Connection\\ for Canonical Gravity\footnote{This
paper is a contribution to the Special Issue ``Loop Quantum Gravity and Cosmology''. The full collection is available at \href{http://www.emis.de/journals/SIGMA/LQGC.html}{http://www.emis.de/journals/SIGMA/LQGC.html}}}

\AuthorNameForHeading{M. Geiller, M. Lachi\`eze-Rey, K. Noui and F. Sardelli}

\Author{Marc GEILLER~$^\dag$, Marc LACHI\`EZE-REY~$^\dag$, Karim NOUI~$^\ddag$ and Francesco SARDELLI~$^\ddag$}
\Address{$^\dag$~Laboratoire APC, Universit\'e Paris Diderot Paris 7, 75013 Paris, France}
\EmailD{\href{mailto:mgeiller@apc.univ-paris7.fr}{mgeiller@apc.univ-paris7.fr}, \href{mailto:mlr@apc.univ-paris7.fr}{mlr@apc.univ-paris7.fr}}

\Address{$^\ddag$~LMPT, Universit\'e Fran\c cois Rabelais, Parc de Grandmont, 37200 Tours, France}
\EmailD{\href{mailto:karim.noui@lmpt.univ-tours.fr}{karim.noui@lmpt.univ-tours.fr}, \href{mailto:francesco.sardelli@lmpt.univ-tours.fr}{francesco.sardelli@lmpt.univ-tours.fr}}

\ArticleDates{Received May 27, 2011, in f\/inal form August 20, 2011;  Published online August 24, 2011}

\Abstract{We construct a Lorentz-covariant connection in the context of f\/irst order cano\-ni\-cal gravity with non-vanishing Barbero--Immirzi parameter. To do so, we start with the phase space formulation derived from the canonical analysis of the Holst action in which the second class constraints have been solved explicitly. This allows us to avoid  the use of Dirac brackets. In this context, we show that there is a ``unique'' Lorentz-covariant connection which is commutative in the sense of the Poisson bracket, and which furthermore agrees with the connection found by Alexandrov using the Dirac bracket. This result opens a new way toward the understanding of Lorentz-covariant loop quantum gravity.}


\Classification{83C05; 83C45}

\section{Introduction}

Loop quantum gravity was founded on the observation by Ashtekar \cite{ashtekar-variables} that working only with the self-dual part of the Hilbert--Palatini action leads to a canonical theory of gravity which is free of second class constraints and features a polynomial Hamiltonian. In Ashtekar's original formulation, the phase space variables are a complex $\su(2)$ connection and a densitized triad f\/ield. Because of the
vector space isomorphism between the complex $\su(2)$ algebra (which is the self-dual or anti self-dual copy of the complex Lorentz
algebra~$\so_{\mathbb{C}}(3,1)$) and  the Lorentz algebra~$\so(3,1)$ itself, working only with the complex~$\su(2)$ connection enables  to preserve the Lorentz covariance.  The drawback of this approach is that working with complex variables requires the imposition of reality conditions in order to recover  the phase space of real general relativity. Of course, if one imposes the reality conditions at the classical level, prior to quantization, one loses all the beauty of the Ashtekar formulation, and recovers the standard Palatini formulation of general relativity, which we do not know how to quantize. Unfortunately, no one knows how to go the other way around, and implement the reality conditions after quantization of the Ashtekar theory.
This dif\/f\/iculty motivated the work of Barbero~\cite{barbero} and, later on, Immirzi~\cite{immirzi}, who introduced a family of canonical transformations, parametrized by the so-called Barbero--Immirzi parameter~$\gamma$, and leading to a canonical theory in terms of a real $\su(2)$ connection known as the Ashtekar--Barbero connection. The action that leads to this canonical formulation was f\/inally found by Holst~\cite{holst}.

Although it is non-physical at the classical level, which is to be expected when performing canonical transformations, the Barbero--Immirzi parameter $\gamma$ does not disappear in the quantum theory: it shows up in the spectra of geometric operators \cite{ashtekar-lewandowski,rovelli-smolin} and in the black hole entropy formula \cite{agullo,ABK,rovelli-black hole, meissner}. This puzzle has led people to question the origin and the physical relevance of this free parameter \cite{rovelli-thiemann}.

Another puzzling feature is that the Ashtekar--Barbero connection is not a Lorentz connection. It is covariant only
under $\su(2)$ gauge transformations. In fact, the notion of Lorentz covariance does not even make sense in this framework.
The reason becomes clear when one looks at the canonical analysis of the Holst action. Indeed,
one uses a partial gauge f\/ixing in the Holst action in order to derive a canonical theory in terms of the Ashtekar--Barbero
connection which is free of second class constraints.
This choice of gauge (which is refered to as the time gauge) corresponds to f\/ixing the direction of the normal to three-dimensional spatial hypersurfaces along the direction of the time axis of the $3+1$ decomposition. By doing so, the $\so(3,1)$ gauge algebra in the internal space is reduced to its rotational $\so(3)$ subalgebra, and the manifest Lorentz covariance of the theory is broken. In other words, the information about how the phase space variables transform under Lorentz boosts is lost. Now if one performs the canonical analysis of the Holst action without this gauge choice, second class constraints appear simply because the connection has more components than the tetrad f\/ield. The appearance of second class constraints makes the classical analysis and then the quantization of the theory much more involved.

In the analysis of constrained systems, there are two ways of dealing with second class constraints: one can either solve them explicitly, or implement them in the symplectic structure by working with the Dirac bracket~\cite{henneaux-teitelboim}.
These two methods are totally equivalent.
Using the Dirac bracket, Alexandrov and collaborators were able to construct a two-parameter family
of Lorentz-covariant connections (which are diagonal under the action of the area operator, and transform properly under the action of spatial dif\/feomorphisms). Generically,
these connections are non-commutative and therefore the theory becomes very dif\/f\/icult to quantize.
However, Alexandrov has opened a path towards a covariant formulation of loop quantum gravity, and has
made the claim that the spectrum of the area operator of the resulting quantum theory should not
depend on the Barbero--Immirzi parameter anymore \cite{alexandrov1,alexandrov2,alexandrov3,alexandrov4}. Furthermore, he has argued that using a covariant connection, instead of the usual $\su(2)$ Ashtekar--Barbero connection, would make the link between the canonical and covariant (spin-foam) quantizations of general relativity much more transparent. However, because of the (generic) non-commutativity of the covariant connection, the quantization of the theory becomes really dif\/f\/icult already at the kinematical level.

The alternative route to deal with covariant connections
was initiated by Barros e S\'a in~\cite{barros}, where, following the idea of Peld\'an~\cite{peldan},
 he showed that it is possible to solve explicitly the second class constraints.
In this approach, the phase space is parametrized by two pairs of canonical variables:
the generalization $(A,E)$ of the usual Ashtekar--Barbero connection and conjugate densitized triad~$E$;
and a new pair of canonically conjugated f\/ields $(\chi,\zeta)$, where $\chi$ and $\zeta$ both take values in $\mathbb R^3$. Then,
Barros e S\'a expressed the remaining boost, rotation, dif\/feomorphism and scalar constraints in terms of these variables. The elegance of this approach is that it enables one to have a simple symplectic structure with commutative variables, and a~tractable expression for the boost, rotation and dif\/feomorphism generators. Although the scalar constraint becomes more complicated, this structure is enough to investigate the kinematical structure of loop quantum gravity in the presence of the full Lorentz group. However, as pointed out in~\cite{alexandrov4}, the drawback is that the Lorentz covariance is apparently lost when the second class constraints are solved. Indeed, there is no obvious Lorentz-covariant connection in the theory. In this context, the present work addresses the following question. Is it possible to f\/ind a~spatial connection that transforms covariantly under the action of the boost and rotation gauge generators? We show that the answer is in the af\/f\/irmative. Besides, we also show that there is a unique Lorentz-covariant connection depending only upon the Ashtekar variable~$A$ and the f\/ield $\chi$. This assumption has a very natural physical interpretation, and has the great advantage of making the connection commutative. Furthermore, this unique connection should coincide with the commutative Lorentz connection studied in~\cite{alexandrov4}, and therefore we could establish a link between the work of Alexandrov
and that of Barros e S\'a. This is precisely what we do in \cite{GLN}. Notice that an earlier attempt to constructing a Lorentz-covariant connection after solving the second class constraints can be found in~\cite{cianfrani-montani}. However, in this paper, the authors use a parametrization of the physical phase space which is dif\/ferent from the one we use in the present work, and their construction is less transparent. We shall therefore argue that the parametrization of Barros e S\'a~\cite{barros} is the most natural one to discuss Lorentz-covariant canonical gravity.

The outline of this work is the following. In the f\/irst section, we brief\/ly review the main results of the Hamiltonian analysis of
the Hilbert--Palatini action with Barbero--Immirzi parameter and without any gauge choice. In particular, we recall the symplectic structure
in terms of the variables solving the second class constraints, and the expression of the boost, rotation and dif\/feomorphism constraints.
In the second section, we outline the proof of the main statement of this letter: there is a unique Lorentz-covariant connection
built with the Ashtekar--Barbero connection $A$ and the f\/ield $\chi$. The technical details of the proof will be given in~\cite{GLN}. Finally, in the last section, we exhibit the constraint that generates spatial dif\/feomorphisms, and show that it acts properly on the Lorentz connection.

Notations are such that $\alpha,\beta,\dots$ refer to spacetime indices, $a,b,\dots$ to spatial indices, $I,J,\dots$ to $\so(3,1)$ indices, and $i,j,\dots$ to $\su(2)$ indices. We will assume that the four-dimensional spacetime manifold $\mathcal{M}$ is topologically $\Sigma\times\mathbb{R}$, where $\Sigma$ is a three-dimensional manifold without boundaries.

\section{A brief analysis of f\/irst order gravity}

 The starting point to study Lorentz-covariant gravity is the generalized Hilbert--Palatini f\/irst order action, with non-trivial Barbero--Immirzi parameter $\gamma$. In terms of the co-tetrad $e=e_\alpha dx^\alpha$ and the Lorentz connection one-form $\omega=\omega_\alpha dx^\alpha$, this action is given by
\begin{gather}\label{Holst}
S[e,\omega]=\int_\mathcal{M}e^I\wedge e^J\wedge\star\gam F[\omega]_{IJ},
\end{gather}
where $F[\omega]_{IJ}$ is the curvature two-form associated with the connection $\omega$ evaluated in the fundamental representation of $\so(3,1)$, and $\star$ denotes the usual Hodge operator in the Lie algebra~$\so(3,1)$. Furthermore, to each vector $\xi\in\so(3,1)$
we associate the element
\[
\gam\xi=(1-\gamma^{-1}\star)\xi\in\so(3,1)  .
\]

The f\/irst order formulation of gravity is appealing mainly because the Lagrangian is polynomial in the fundamental variables~$e$ and~$\omega$. Therefore, one might a priori expect the quantization of gravity to be simpler in the f\/irst order formulation than in the standard second order metric formulation. However, this viewpoint is a bit naive. Indeed, to make the Lagrangian polynomial in the variables $e$ and $\omega$, one adds extra non-physical degrees of freedom at each space-time point, and, apart from the case of three-dimensional gravity, these extra degrees of freedom are not in general pure gauge, and have to be identif\/ied and eliminated prior to quantization. After this procedure, one ends up  again with standard second order gravity~\cite{peldan}, and comes to the conclusion that f\/irst order gravity is not a particularly interesting starting point for the quantization of general relativity.

However, Ashtekar realized~\cite{ashtekar-variables} that working only with the self-dual (or anti self-dual) part of the connection was enough to eliminate all these extra non pure gauge degrees of freedom. The resulting Ashtekar action for gravity, which can be obtained by choosing $\gamma^2=-1$ (in the Lorentzian case) in the generalized Holst action~(\ref{Holst}), is polynomial in the fundamental variables. However, since it is formulated in terms of complex variables, one has to impose reality conditions through a complicated non-polynomial equation that is still poorly understood. As we mentioned in the introduction, this dif\/f\/iculty motivated the work of Barbero~\cite{barbero} and Immirzi~\cite{immirzi}, who introduced a family of canonical transformations, parametrized by the so-called Barbero--Immirzi parameter~$\gamma$, and leading to a theory of real gravity without extra non pure gauge degrees of freedom. In this approach, the fundamental variable is the so-called Ashtekar--Barbero connection, all the phase space variables are real, and the Hamiltonian constraint is slightly more complicated (and non-polynomial) than in the case of Ashtekar complex gravity. The Ashtekar--Barbero connection is the starting point for canonical loop quantum gravity, and allows to build a complete and consistent description of the kinematical Hilbert space. However, this construction is valid in the time gauge only, where the Ashtekar--Barbero connection is an $\su(2)$ connection and not a fully Lorentz-covariant connection~\cite{samuel}. Many properties of the kinematical Hilbert space, like the spectra of geometric operators, are based on the fact that the connection takes values in~$\su(2)$ and not in~$\so(3,1)$.

This observation naturally led Alexandrov and collaborators~\cite{alexandrov1,alexandrov2,alexandrov3,alexandrov4} to ask whether the properties of the kinematical Hilbert space remain unchanged if we do not work in the time gauge. To clarify this issue, they started with the generalized Holst action~(\ref{Holst}), and performed its canonical analysis without this gauge choice. It is immediate to see that the 16 components of the co-tetrad f\/ield $e_\alpha^I$ contain 4 Lagrange multipliers ($e^I_0$), and that the 24 components of the $\so(3,1)$ spin connection $\omega^{IJ}_\alpha$ contain 6 Lagrange multipliers ($\omega^{IJ}_0$). These multipliers impose the constraints
\begin{gather}\label{constraints}
\mathcal{H}=\pi^a_{IK}\pi^{bK}_J\gam F^{IJ}_{ab},\qquad
\mathcal{H}_a=\pi^b_{IJ}\gam F^{IJ}_{ab},\qquad
\mathcal{G}_{IJ}=D_a\gam\pi^a_{IJ},
\end{gather}
which are expressed in terms of the spatial components of the connection~$\omega_a^{IJ}$ and the  momenta~$\pi^a_{IJ}$. These momenta are
def\/ined as the derivative of the  Lagrangian $\mathcal{L}[e,\omega]$ with respect to $\dot{\omega}_a^{IJ}\equiv\partial_0 {\omega}_a^{IJ}$,
and are explicitly given in term of the co-tetrad f\/ield by
\[
 \pi^a_{IJ}=\f{\delta\mathcal{L}}{\delta\dot{\omega}^{IJ}_a}=\epsilon_{IJKL}\epsilon^{abc}e^K_be^L_c.
\]
In the constraints (\ref{constraints}), we have introduced the notation $D$ for the covariant derivative associated with the Lorentz connection~$\omega$.

At this stage of the analysis, the Hamiltonian theory is described by the 18 components $\omega_a^{IJ}$ of the $\so(3,1)$ connection, and the 12 components $e_a^I$ of the co-tetrad f\/ield, constrained to satisfy the 10 relations (\ref{constraints}).
It is therefore straightforward to realize that in order to recover the 4 phase space degrees of freedom per spatial point, we have to eliminate the extra non pure gauge phase space variables. In the language of constrained systems, this means that the theory admits secondary constraints, which in addition have to be second class. This is indeed the case. To understand how this comes about, it is simpler to work with the momenta $\pi^a_{IJ}$ instead of the co-tetrad components $e^I_a$, and, as it was initially suggested in \cite{peldan}, add the simplicity condition
\[
\mathcal{C}^{ab}\equiv\star\pi^a_{IJ}\pi^{bIJ}\approx0.
\]
Classically, it is equivalent to work with the 12 components $e_a^I$, or with the 18 components $\pi^a_{IJ}$ constrained by the 6 equations $\mathcal{C}^{ab}\approx0$. As a consequence of this procedure, the non-physical Hamiltonian phase space is now parametrized by the pairs of canonically conjugated variables $(\omega_a^{IJ},\pi^a_{IJ})$, with the set of constraints (\ref{constraints}) to which we add  $\mathcal{C}^{ab}\approx0$. These new constraints generate 6 additional secondary constraints{\samepage
\[
\mathcal{D}^{ab}=\star\pi^c_{IJ}\big(\pi^{aIK}D_c\pi^{bJ}_{~~K}+\pi^{bIK}D_c\pi^{aJ}_{~~K}\big)\approx0,
\]
and the Dirac algorithm closes \cite{peldan,alexandrov2,barros}.}

Then, one has to make the separation between f\/irst class and second class constraints. To do so, one computes the Dirac matrix, whose components are given by the Poisson brackets $\big\{\phi_1,\phi_2\big\}$ between any pair of constraints $(\phi_1,\phi_2)$. The dimension of this matrix is $22\times22$, but it is rather immediate to show that its kernel is 10-dimensional. Physically, this means that among the 22 constraints $\mathcal{H}$, $\mathcal{H}_a$, $\mathcal{G}_{IJ}$, $\mathcal{C}^{ab}$ and $\mathcal{D}^{ab}$, 10 are f\/irst class, and the remaining 12 are second class. Moreover, one can check that $\mathcal{C}^{ab}\approx0$ and $\mathcal{D}^{ab}\approx0$ form a set of second class constraints. Since the constraints (\ref{constraints}) are f\/irst class (up to the previous second class constraints), they generate the symmetries of the theory, namely the space-time dif\/feomorphisms and the Lorentz gauge symmetry. Finally, we end up with a phase space parametrized by 18 connection components and 18 conjugate momenta, with the 10 f\/irst class constraints $\mathcal{H}$, $\mathcal{H}_a$ and $\mathcal{G}_{IJ}$, and the 12 second class constraints $\mathcal{C}^{ab}$ and $\mathcal{D}^{ab}$. As expected, this leads to 4 degrees of freedom per spatial point in the phase space.

However, the canonical analysis of the Holst action does not end with this counting of degrees of freedom. One has to ``eliminate'' the second class constraints, either by solving them explicitly, or by computing the Dirac bracket that they def\/ine. In order to keep the Lorentz covariance manifest, Alexandrov and collaborators have chosen to use the Dirac bracket, which led to a quite complicated description of the kinematical phase space. In particular, the Lorentz-covariant connection that they found has a very complicated expression, and is generically non-commutative in the sense of the Dirac bracket. As a result, no one knows how to quantize this theory for the moment~\dots.

Following the idea of \cite{peldan}, Barros e S\'a has proposed a solution to the second class constraints. His approach is rather
elegant because it results in a parametrization of the phase space with the two pairs of canonical variables
\begin{gather}\label{phasespace}
\big\{A^i_a(x),E^b_j(y)\big\}=\delta^i_j\delta^b_a\delta^3(x-y),\qquad
\big\{\chi_i(x),\zeta^j(y)\big\}=\delta^j_i\delta^3(x-y),
\end{gather}
where
\begin{gather}\label{A}
A^i_a\equiv\gam\omega^{0i}_a+\gam\omega^{ij}_a\chi_j,
\end{gather}
together with the usual scalar, dif\/feomorphism and Lorentz constraints. In particular, the full Lorentz gauge generator $\mathcal{G}_{IJ}$ can be split between its boost part $\mathcal{B}_i\equiv\mathcal{G}_{0i}$, and its rotational part $\mathcal{R}_i\equiv(1/2)\epsilon_i^{~jk}\mathcal{G}_{jk}$.

The physical interpretation of the canonical variables (\ref{phasespace}) is clear: $A$ is a generalization of the Ashtekar--Barbero connection when the time gauge is not imposed, $E$ is the canonical electric f\/ield, $\chi$ encodes the deviation of the normal to the hypersurfaces from the time direction, and~$\zeta$ is its conjugate momentum.

With this formulation, we have a parametrization of the kinematical phase space which is free of second class constraints, but
the Lorentz covariance remains somehow hidden (because it seems that the $\su(2)$ connection $A$ is given a special status). Moreover, the expression of the scalar constraint is very complicated already at the classical level. Nonetheless, we are going to show in the following section that we can restore the Lorentz covariance, and that, under some natural hypothesis, it is possible to f\/ind a unique Lorentz-covariant connection. One of the reasons for this is that the expression of the boost and rotational constraints is rather simple. Indeed, in terms of the phase space variables (\ref{phasespace}), the smeared constraints
\[
\mathcal{B}(u)=\int d^3x\,\mathcal{B}_i(x)u^i(x),\qquad \mathcal{R}(v)=\int d^3x\,\mathcal{R}_i(x)v^i(x),
\]
where $u$ and $v$ are vectors in $\mathbb R^3$, simply read
\begin{gather*}
\mathcal{B}(u) = \int d^3x\left[-\partial_au\cdot\left(E^a-\f{1}{\beta}\chi\wedge E^a\right)-u\cdot(\chi\wedge E^a)\wedge A_a-u\cdot\zeta+(\zeta\cdot\chi)(\chi\cdot u)\right], \\
\mathcal{R}(v) = \int d^3x\left[\partial_av\cdot\left(\chi\wedge E^a+\f{1}{\beta}E^a\right)+v\cdot(A_a\wedge E^a-\zeta\wedge\chi)\right].
\end{gather*}
Here we have identif\/ied $A_a^i$ and $E^a_i$ with the components of the vectors $A_a$ and $E^a$ in $\mathbb R^3$, and the notation $x\wedge y$ denotes the standard wedge product between two vectors $x$ and $y$ in $\mathbb R^3$. Note that indices in $\mathbb R^3$ are raised and lowered with the Euclidean f\/lat metric $\text{diag}(+1,+1,+1)$. Starting from the Poisson brackets~(\ref{phasespace}), it is long but straightforward to verify that these boost and rotational constraints satisfy the Lorentz algebra
\begin{gather*}
\big\{\mathcal{B}(u_1),\mathcal{B}(u_2)\big\} = -\mathcal{R}(u_1\wedge u_2), \\
\big\{\mathcal{R}(v_1),\mathcal{R}(v_2)\big\} = \mathcal{R}(v_1\wedge v_2), \\
\big\{\mathcal{B}(u_1),\mathcal{R}(v_2)\big\} = \mathcal{B}(u_1\wedge v_2),
\end{gather*}
for any $u_i$ and $v_i$. For the sake of convenience, we will use the notation $\mathcal{G}(\xi)=\mathcal{B}(u)+\mathcal{R}(v)$ to denote the full Lorentz generator, with $\xi=(u,v)$, where $u$ is the boost part of $\xi$ and $v$ its rotational part.

Now we have a simple symplectic structure (\ref{phasespace}) on the phase space, and the expression of the generators of the Lorentz algebra. This is all we need to construct a fully Lorentz-covariant connection, which is what we do in the next section.

Before going into further details, let us f\/inish this section by quickly analyzing what happens when the time gauge is imposed. To work with the time gauge, we impose the condition $\chi\approx0$, which has to be interpreted as a new constraint in the theory. It is clear that this constraint, together with the boost generator, form a set of second class constraints, that we can solve explicitly by taking $\chi=0$ and $\zeta=\partial_aE^a$. By doing so, the variables $(\chi,\zeta)$ are eliminated from the theory, and we see from the expression (\ref{A}) that $A$ becomes the standard Ashtekar--Barbero connection. In other words, we recover the usual $\su(2)$ theory.

\section{The unique Lorentz-covariant connection}

 Our strategy is based on the idea that the Lorentz-covariant connection $\mathcal{A}$ that we are looking for depends only on the variables $A$ and $\chi$. The motivation for this is twofold. First, we would like to work with a commutative connection in order to have a chance to construct at least the kinematical Hilbert space following the techniques of usual loop quantum gravity. Secondly, we would like to interpret $\chi$ as a boost generator which sends the Lorentz-covariant connection $\mathcal{A}$ to an $\su(2)$ time gauge connection \cite{GLN}. If this is indeed the case, then $\mathcal{A}$ will naturally depends on $A$ and $\chi$ solely.

Now, we address the following question. Can we construct an $\so(3,1)$-valued one-form $\mathcal{A}$, depending only on the variables $A$ and $\chi$, such that
\begin{gather}\label{covarianceeq}
\big\{\mathcal{G}(\xi),\mathcal{A}\big\}=d\xi+[\mathcal{A},\xi],
\end{gather}
where $\xi\in\mathbb R^6$ is identif\/ied with an element $(u,v)\in\so(3,1)$?
If such a one-form can be found, it can clearly be interpreted as a Lorentz-covariant connection. To answer this question, we proceed in three steps:
\begin{enumerate}\itemsep=0pt
\item[1)] f\/irst, we compute how the variables $A$ and $\chi$ transform under the action of the Lorentz generator $\mathcal{G}(\xi)$;

\item[2)] then, we f\/ind the general form of the connection $\mathcal{A}$, assuming that it depends only on the variables $A$ and $\chi$;

\item[3)] f\/inally, we look at the solutions to equation (\ref{covarianceeq}).
\end{enumerate}

\subsection[Lorentz transformations of $A$ and $\chi$]{Lorentz transformations of $\boldsymbol{A}$ and $\boldsymbol{\chi}$}

 The action of the boosts $\mathcal{B}(u)$ and rotations $\mathcal{R}(v)$ on the variables $A$ and $\chi$ is given by
\begin{gather*}
\big\{\mathcal{B}(u),A\big\} = du-\f{1}{\gamma}du\wedge\chi-(u\wedge A)\wedge\chi, \\
\big\{\mathcal{B}(u),\chi\big\}  = u-(u\cdot\chi)\chi, \\
\big\{\mathcal{R}(v),A\big\}  = -dv\wedge\chi-\f{1}{\gamma}dv+A\wedge v, \\
\big\{\mathcal{R}(v),\chi\big\} = \chi\wedge v.
\end{gather*}
We immediately remark that the action of the boosts on the vector~$\chi$ is non-linear, which might lead to some complications later on. Furthermore, the physical interpretation of this transformation is not clear. For these reasons, we introduce the following change of variables (which is in fact part of a canonical transformation on the phase space~(\ref{phasespace})~\cite{GLN}). Assuming that $\chi^2<1$, we def\/ine the new vector $Y=T\chi$,
with $T=(1-\chi^2)^{-1/2}$. Then, it is immediate to see that the action of the Lorentz generators on the variables $T$ and $Y$, which is given by
\[
\big\{\mathcal{B}(u),T\big\}=Y\cdot u,\qquad
\big\{\mathcal{B}(u),Y\big\}=Tu,\qquad
\big\{\mathcal{R}(u),T\big\}=0,\qquad
\big\{\mathcal{R}(u),Y\big\}=Y\wedge v,
\]
coincides with the linear transformation of the four-vector $(T,Y)$ under the action of the inf\/initesimal Lorentz matrix
\[
(T,Y)
\begin{pmatrix}
 0 & {}^{\text{\tiny{T}}}u \\
 u & \vb
\end{pmatrix}
=(Y\cdot u,Tu+Y\wedge v).
\]
Here, we have associated to any vector $v=(v_1,v_2,v_3)\in\mathbb R^3$ the three-dimensional antisymmetric rotational matrix
\[
\underline{v}=
\begin{pmatrix}
 0 & -v_3 & v_2 \\
 v_3 & 0 & -v_1 \\
 -v_2 & v_1 & 0
\end{pmatrix},
\]
which def\/ines an inf\/initesimal rotation around the axis parallel to  $v$. In particular, we have that $\underline{v}u=v\wedge u$ for any pair of vectors $(u,v)\in\mathbb{R}^6$.

From now on, we will therefore consider the vector $Y$ instead of $\chi$ for our calculations, and look for a Lorentz connection that depends on $A$ and $Y$.

\subsection[General form of the Lorentz connection $\cal A$]{General form of the Lorentz connection $\boldsymbol{\cal A}$}

 A Lorentz connection $\mathcal{A}$ is an $\so(3,1)$-valued one-form. It is convenient to decompose it into its boost and rotational components as $\mathcal{A}=\bst\mathcal{A}+\rot\mathcal{A}$, where the notations are transparent. Thanks to the vector space isomorphism $\so(3,1)\simeq\mathbb{R}^3\oplus\mathbb{R}^3$, each of the two components can be identif\/ied with a one-form taking values in $\mathbb R^3$. As a consequence, our problem is now reduced to f\/inding these two $\mathbb R^3$-valued one-forms.

Fortunately, we have two natural $\mathbb{R}^3$-valued one-forms at our disposal: the variable $A$ itself, and the exterior derivative $dY$ of the vector $Y$. As a consequence, the general solution for~$\bst\mathcal{A}$ or~$\rot\mathcal{A}$, generically denoted~$\brA$, is necessarily of the form
\[
\brA=MA+NdY,
\]
where $M$ and $N$ are two $3\times 3$ matrices. The only matrices that we can construct from~$Y$ and~$A$, are~$\Yb$ itself, and any power $\Yb\!^x$, where $x$ can be a priori any real number. However, as one can see from the characteristic polynomial
\begin{gather}\label{polynomial}
\Yb\!^3+Y^2\Yb=0,
\end{gather}
where $Y^2=Y^iY_i$ is the square norm of the vector $Y$, the matrix $\underline{Y}$ is not invertible, and admits purely imaginary non-trivial eigenvalues. This shows that we cannot consider any arbitrary real power $\Yb\!^x$. The exponent $x$ must be non-negative, and also an integer in order to have a~connection with values in $\mathbb R^3$ and not in $\mathbb C^3$. In summary, because of the form of the characteristic polynomial (\ref{polynomial}), it turns out that the most general $\mathbb R^3$-valued one-form can be written as
\begin{gather}\label{generalform}
\brA=\left(a_0+a_1\Yb+a_2\Yb\!^2\right)A+\left(b_0+b_1\Yb+b_2\Yb\!^2\right)dY,
\end{gather}
where $a_i$ and $b_i$ are a priori functions of the coordinates $Y_i$ only.

As a consequence, we can now look for a Lorentz-covariant connection $\mathcal{A}$ admitting a pure boost component $\bst\mathcal{A}$, and a pure rotational component $\rot\mathcal{A}$, both being of the form (\ref{generalform}).

\subsection{Solution to the Lorentz covariance condition}

 To f\/ind the solutions to (\ref{covarianceeq}), we study separately the transformation of the connection $\mathcal{A}$ under the boosts and the rotations. The transformations are given by
\begin{gather*}
\big\{\mathcal{R}(v),\mathcal{A}\big\} = dv-\vb\!\left(\rot\mathcal{A}+\bst\mathcal{A}\right), \\
\big\{\mathcal{B}(u),\mathcal{A}\big\} = du-\ub\!\left(\rot\mathcal{A}-\bst\mathcal{A}\right) .
\end{gather*}

It is straightforward to notice that in order for $\mathcal{A}$ to transform properly under the action of~$\mathcal{R}(v)$, the coef\/f\/icients~$a_i$ and~$b_i$ must be invariant under the action of the rotations. Therefore, they can only depend on the variable~$T$ (or equivalently $Y^2$). Moreover, these coef\/f\/icients have to satisfy a set of additional algebraic relations that we do not detail here~\cite{GLN}.

Then, the requirement of covariance under the action of the boosts is the condition that uniquely determines the connection. It f\/ixes completely all the 12 coef\/f\/icients $a_i$ and $b_i$, which depend only on $T$ and $\gamma$. We f\/ind that the boost and rotation components are given by
\begin{gather*}
\bst\mathcal{A} = \gamma T \Yb A+\f{1}{T}\left(1-\gamma T \Yb-\Yb\!^2\right)dY, \\
\rot\mathcal{A} = -\gamma\left(1-\Yb\!^2\right)A+\f{\gamma}{T}\left(1+\f{T}{\gamma} \Yb-\Yb\!^2\right)dY.
\end{gather*}
The Lorentz-covariant connection  can be expressed in a simpler form if we replace the variable~$Y$
by its expression in terms of the original vector $\chi$. Indeed, a simple calculation leads to
\begin{gather}
\bst\mathcal{A}=\f{1}{1-\chi^2}\left(d\chi+\Omega_0\wedge\chi\right),\qquad
\rot\mathcal{A}=\f{1}{1-\chi^2}\left(\chi\wedge d\chi+\Omega_0\right),\label{uc}
\end{gather}
where $\Omega_0=\gamma\left(d\chi-A+(A\cdot\chi)\chi\right)$.

\looseness=-1
It is remarkable that this unique connection transforms also correctly under the action of the rotations. As a consequence, we end up with a unique Lorentz-covariant connection, with boost and rotational components respectively given by (\ref{uc}). In that sense, we claim that there is a unique commutative Lorentz-covariant connection in f\/irst order general relativity. Obviously, this connection reduces to the standard Ashtekar--Barbero
connection when the time gauge is imposed.

\subsection{Action of spatial dif\/feomorphisms}

 As usual, the generator of spatial dif\/feomorphisms turns out to be given by a linear combination of the vector and the Gauss constraints. In the present context, it is given by the following combination of the vector constraint $\mathcal{H}_a$, and the generators $\mathcal{B}$ and $\mathcal{R}$ of the Lorentz algebra:
\begin{gather*}
\tilde{\mathcal{H}}_a(N^a) = \mathcal{H}_a(N^a)-\f{\gamma}{1+\gamma^2}\big(\gamma\mathcal{B}(N^aA_a)-\mathcal{R}(N^aA_a)\big) \\
 \phantom{\tilde{\mathcal{H}}_a(N^a)}{} = \int d^3x\,N^a\big(E^b\cdot\partial_aA_b+\zeta\cdot\partial_a\chi-E^b\cdot\partial_bA_a-A_a\cdot\partial_bE^b\big).\nonumber
\end{gather*}
Indeed, $\tilde{\mathcal{H}}_a(N^a)$ generates the transformations
\begin{gather*}
\big\{\tilde{\mathcal{H}}_a(N^a),A_b\big\} = -N^a\partial_aA_b-A_a\partial_bN^a=-\pounds_{N^a}A_b, \\
\big\{\tilde{\mathcal{H}}_a(N^a),\chi\big\} = -N^a\partial_a\chi=-\pounds_{N^a}\chi,
\end{gather*}
where $\pounds_{N^a}$ denotes the Lie derivative along the vector f\/ield $N^a$. We see that the phase-space variables $A$ and $\chi$ transform respectively as a one-form and a scalar. It is therefore clear that the connection $\mathcal{A}$ will also transform properly under the action of dif\/feomorphisms of the spatial hypersurface.

\vspace{-2mm}

\section{Discussion and outlook}

 Starting from the canonical analysis of the non-time gauge Holst action with non-vanishing Barbero--Immirzi parameter, carried out by Barros e S\'a in \cite{barros}, we have constructed a commutative (in the sense of the Poisson bracket) Lorentz-covariant connection. In this framework, where the second class constraints are explicitly solved, it is not necessary to consider the Dirac bracket anymore. In that sense, it of\/fers an alternative to the work of Alexandrov on Lorentz-covariant canonical gravity~\cite{alexandrov1,alexandrov2,alexandrov3,alexandrov4}. This approach has never been considered previously because solving the second class constraints seems to ``hide'' the Lorentz covariance. We have shown how it can be restored in a unique fashion, by constructing a connection that has a simple expression in terms of the canonical variables on the phase space.

Even if this result has already been derived in another framework, our approach opens a~new way toward the understanding of Lorentz-covariant loop quantum gravity. Indeed, it has the advantage of being technically much simpler and transparent. It is based on a natural enlargement of the usual Ashtekar--Barbero phase space with the addition of a single pair of canonical variables $(\chi,\zeta)$. The constraint algebra is similar to that of usual loop quantum gravity, with only an extra constraint corresponding to the boost transformations, and a slightly more complicated expression.
We hope that this structure will enable us to clarify the geometrical interpretation of the results obtained previously by Alexandrov. It is going to serve as a starting point for the construction of the kinematical Hilbert space of the quantum theory and that of the basic geometric operators.

Furthermore, we hope that the present framework will shed some new lights on the relation between the canonical and the spin-foam quantizations of general relativity. The usual point of view is that the $\SL(2,\mathbb{C})$ spin-foam dynamics in the bulk def\/ines $\SU(2)$ boundary states which span a subset of the projected spin network states \cite{rovelli-speziale}. Such states are $\SL(2,\mathbb{C})$ functionals which are entirely characterized by their restriction to $\SU(2)$. Since Lorentz-covariant canonical loop quantum gravity enables one to work directly with a boundary $\SL(2,\mathbb{C})$ theory, it would be interesting to clarify the relationship between this kinematical structure and the various proposals for the spin-foam dynamics which exist in the literature~\cite{EPRL,FK}.

But before starting this enterprise, we have to understand the classical properties of this Lorentz-covariant connection. In particular, we have to study its behavior under the action of the time dif\/feomorphisms. Also, we have indications that this Lorentz connection might be related to the connection found by Alexandrov. Furthermore, we think that it is possible to recover with our approach the whole family of Lorentz connections found by Alexandrov, and we hope that this will clarify the issue of non-commutativity. Finally, it would be interesting to study if the connection $\mathcal{A}$ is conjugated to an $\su(2)$ connection under the action of a boost. This last property would be very useful for the construction of the spin-network representation and the study of the geometric operators.
These questions are addressed in~\cite{GLN}, where we study the quantization of the Lorentz-covariant theory.

\subsection*{Acknowledgements}

The authors would like to thank S.~Alexandrov for carefully reading an earlier draft of this work, and for useful comments. K.N.\ is partially supported by the ANR.
\vspace{-3mm}

\pdfbookmark[1]{References}{ref}
\LastPageEnding

\end{document}